# ENHANCED ARCHITECTURES FOR ROOM-TEMPERATURE REVERSIBLE LOGIC GATES IN GRAPHENE

Daniela Dragoman[1] and Mircea Dragoman[2]

[1]Univ. Bucharest, Physics Faculty, P.O. Box MG-11, 077125 Bucharest, Romania

[2] National Institute for Research and Development in Microtechnology (IMT), P.O. Box 38-160, 023573 Bucharest, Romania

## Abstract

We show that reversible two- and three-input logic gates, among which we mention the universal Toffoli gate, can be implemented with three tilted gating electrodes patterned on a monolayer graphene flake. These low-dissipation gates are based on the unique properties of ballistic charge carriers in graphene, which induce the appearance of bandgaps in transmission for properly potential barriers. The enhanced architectures for reversible logic gate implementations proposed in this paper offer a remarkable design simplification in comparison to standard approaches based on field-effect transistor circuits and a potential high-frequency operation.

## 1. Introduction

The quest for reversible logic gates, i.e. logic gates that implement a bijective relation between input and output states, is motivated by their assumed low energy dissipation [1]. The present day Boolean computers are based on irreversible logic operations, generally implemented by field-effect-transistors (FETs) fabricated in CMOS technology [2]. These computers, whether attaining a very large scale integration, of about $10^9$ transistors per chip, dissipate a large amount of heat, which limits further development of an even more enhanced integration. Reversible computing, which originally attracted interest in the context of the inherently reversible quantum computing [3], could solve this bottleneck. However, reversible logic gates could function also in the context of classical computers, dissipating much less energy than irreversible Boolean gates.

Up to now, there were several attempts to implement reversible logic gates using CMOS technology, but a relatively large number of FETs, with associated interconnections, were required for the realization of even simple reversible gates [4]. In this paper we propose an enhanced architecture for the implementation of reversible logic gates based on the unique properties of ballistic electrons in monolayer graphene [5]. In particular, we use the negative differential resistance (NDR) of a simple potential barrier in graphene, which can be induced by a tilted electrode that acts as electrostatic gate [6]. By generalizing this principle to several tilted electrodes, it is possible to implement reversible logic gates, such as CNOT and the universal Toffoli gate, in simplified architectures that could work at THz frequencies. This radical approach for implementing reversible logic gates differs entirely from the present trend in classical graphene-based gates, which focuses on graphene FETs for basic irreversible logic gates [7-13]. Due to the lack of bandgap in graphene, very often these graphene logic gates function optimally either at low temperatures [8,9] or in special environments [10], which make them difficult to compete with CMOS technology, despite the high mobility of electrons in graphene. Our proposed implementation of reversible logic gates working at ambient conditions and using simplified architectures could become a possible alternative to CMOS technology. In this respect, very recently ballistic devices at the wafer scale comprising tilted gating electrodes on CVD graphene monolayer were demonstrated to function at ambient conditions [14]. Although the implementation of reversible logic gates does not mean yet a reversible computer, it is the first step towards this direction, which would require a re-thinking of the way in which logic operations/algorithms are performed.

## 2. Implementation of logic gates in graphene

The configuration that could implement reversible logic gates in graphene based on the NDR effect is presented in Fig. 1. The tilted electrodes, which could tune the electrostatic potential energy of electrons to desired values $U_{g1}$, $U_{g2}$ and $U_{g3}$ via suitable applying voltages, are used to define the logic inputs, 0 and 1. The output logic states are encoded in the current value measured between the outer contacts. With the device in Fig. 1, it is possible to implement one-input logic gates (for example, NOT) when only one gating electrode is active (i.e., a voltage is applied on it), two-input logic gates if voltages are applied on two electrodes, and three-input logic gates if all electrodes are active. The dimension of the ballistic device is limited by the mean-free path of charge carriers, which reaches 400 nm in graphene monolayers at room temperature, and even 1 μm in graphene deposited over boron nitride [15]. Because the minimum feature of nanolithography techniques is 5 nm and even less, several tilted electrodes or cascaded logic gates can be fabricated on a graphene flake, the implementation of more sophisticated algorithms comprising reversible gates becoming thus possible. The computation of transmission coefficients in this paper are performed using transfer matrices for ballistic propagation in a succession of graphene regions with different potential energies [16], and the currents are then estimated using the Landauer formula for a Fermi energy equal to zero.

## 2.1. Implementation of two-input logic gates

Two-input logic gates include irreversible logic gates such as OR and AND, as well as reversible logic gates, such as CNOT. All these gates can be implemented using the first two tilted electrodes in the configuration depicted in Fig. 1.

For instance, an OR gate, the truth table of which is given in Table 1, can be implemented if the input 0 logic state of the first bit, $B_1$, corresponds to no applied voltage on the first tilted electrode and the input 1 logic state is associated with a voltage that shifts the potential energy in the region beneath this electrode with $U_{g1}$. In this case, for $\varphi = 15^o$, $U_{g1} = 0.2$ eV and $d_1 = 30$ nm, the transmission and current are represented with blue dashed lines in Fig. 2 for the input logic state 0 and with dotted red lines for 1: the current value measured at a bias of $V = 0.45$ V is high for the 0 logic state and low for 1. (Note that if the input states are encoded as before in terms of applied voltages on the electrode, a one-input NOT gate can be implemented by associating to the high-current values at $V = 0.45$ V the logic state 1 and to the low-current values the state 0.) With the same assignment for the second bit, $B_2$, the measured current will have the values designated for a single bit if one of the bits has the logic value 0 (if one of the two electrodes is inactive), and the bias dependence represented with black solid line in Fig. 2 if the input state is 11. The last curve was calculated for $U_{g1} = U_{g2} = 0.2$ eV, $d_1 = d_2 = l_1 = 30$ nm. Therefore, if we associate to the logic state 0 a high current at a bias $V = 0.45$ V, and to 1 a low current value at the same bias, the measured output current encodes the logic states of the OR gate.

An AND gate can be implemented in a similar manner, taking into account that the current at $V = 0.45$ V is high only when no voltages are applied on the two tilted electrodes. Then, a re-assignment of the logic states is needed compared to the OR gate, such that the truth table of the AND gate is fulfilled. The truth table and the assignment of logic states for the AND gate are given in Tables 3 and 4, respectively.

Besides the irreversible OR and AND gates, the device in Fig. 1 can implement a reversible controlled-NOT (CNOT) gate, the table of truth of which is given in Table 5. A two-input CNOT gate applies a NOT operation on a target bit B only if the control bit C has the logical value 1. Defining the logical values 1 of bits B and C as in Table 6, i.e. by applying voltages on the first two tilted electrodes that modify the potential energies with $U_{g1} = 0.2$ eV and $U_{g2} = 0.4$ eV, respectively, and the logical values 0 as the states in the absence of gating voltages, the transmission coefficients and currents for logical values of BC of 00, 10, 01 and 11 are represented in Fig. 3 with blue dashed line, red dotted line, magenta dashed-dotted line and black solid line, respectively. We have assumed in the simulations that $\varphi = 15$ $^{\mathrm{o}}$ and $d_1 = d_2 = l_1 = 30$ nm. The output logical value of the CNOT gate can then be defined in terms of the current value at 0.45 V as indicated in Table 6, i.e. this definition is correlated with the logical state of C. As C is a control bit, its logical value should be known; if it is not known, it can be found by measuring the current value for the bias $V = 0.9$ V. The operation of this CNOT gate is based on the possibility of assigning different logical values to output states depending on the logical state of C; as the assignment is a matter of convention, a (properly defined) flexibility of the assignment procedure could help simplify the implementation of logic gates.

## 2.2. Implementation of three-input logic gates

As a textbook example of a three-input logic gates we choose the Toffoli gate, which is in fact a NOT operation performed on the target bit B if both control bits $C_1$ and $C_2$ are 1 (more precisely, if $C_1$ AND $C_2$ is 1); the truth table of this gate is given in Table 7. A possible implementation of this gate is achieved by defining the 0 and 1 states of B as corresponding to $U_{g1} = 0$ and 0.2 eV, respectively, while the logical 0 and 1 states of the control bits are associated to gate voltages on the last two electrodes that induce $U_{gi} = 0.4$ eV and $U_{gi} = 0$, respectively, $i = 2,3$. Then the logical values of the output of the Toffoli gate can be defined according to Table 8, the current value at a bias of $V = 0.9$ V indicating (if it is not already known) the logical values of $C_1$ AND $C_2$. Note that Toffoli is a universal reversible gate, such that any reversible circuit can be build from a succession of Toffoli gates.

## 3. Conclusions

We have shown that reversible two- and three-input logic gates can be implemented with just three tilted electrodes on a monolayer graphene flake. In particular, the universal Toffoli gate can be implemented in this way. The implementation of these gates is based on the unique bandgapless Dirac-like behavior of ballistic charge carriers in graphene, which leads to the appearance of bandgap in transmission for properly tilted gating electrodes. In comparison to the standard approach of implementing reversible logic gates using circuits consisting of irreversible gates based on FETs, the architecture proposed in this paper offers a remarkable design simplification. Moreover, it involves ballistic electrons in the material with potentially the highest mobility, fact that recommends graphene-based NDR devices for ultrafast reversible and thus, low-dissipation computation. The enhanced architectures presented in this paper could be seen as a first step towards reversible computation based on new concepts that could lead to a real improvement in computation facilities in miniaturized computers.

Table 1: Truth table for OR gate

| $B_1$ | $B_2$ | $B_1$ OR $B_2$ |
|-------|-------|----------------|
| 0 | 0 | 0 |
| 0 | 1 | 1 |
| 1 | 0 | 1 |
| 1 | 1 | 1 |

Table 2: Assignments of logic values for the implementation of OR gate

| | Logic value | Current |
|---|---|---|
| $B_1$ | 0 | High at $V = 0.45$ V |
| | 1 | Low at $V = 0.45$ V |
| $B_2$ | 0 | High at $V = 0.45$ V |
| | 1 | Low at $V = 0.45$ V |
| $B_1$ OR $B_2$ | 0 | High at $V = 0.45$ V |
| | 1 | Low at $V = 0.45$ V |

Table 3: Truth table for AND gate

| B$_1$ | B$_2$ | B$_1$ AND B$_2$ |
|-------|-------|-----------------|
| 0 | 0 | 0 |
| 0 | 1 | 0 |
| 1 | 0 | 0 |
| 1 | 1 | 1 |

Table 4: Assignments of logic values for the implementation of AND gate

|  | Logic value | Current |
|--|-------------|---------|
| B$_1$ | 0 | Low at $V = 0.45$ V |
|  | 1 | High at $V = 0.45$ V |
| B$_2$ | 0 | Low at $V = 0.45$ V |
|  | 1 | High at $V = 0.45$ V |
| B$_1$ AND B$_2$ | 0 | Low at $V = 0.45$ V |
|  | 1 | High at $V = 0.45$ V |

Table 5: Truth table for CNOT gate

| B | C | CNOT |
|---|---|------|
| 0 | 0 | 0 |
| 0 | 1 | 1 |
| 1 | 0 | 1 |
| 1 | 1 | 0 |

Table 6: Assignments of logic values for the implementation of CNOT gate

| | Logic value | Current |
|---|---|---|
| B | 0 | High at $V = 0.45$ V |
| | 1 | Low at $V = 0.45$ V |
| C | 0 | High at $V = 0.9$ V |
| | 1 | Low at $V = 0.9$ V |
| CNOT | 0 | High at $V = 0.45$ V when C = 0 <br><br> Low at $V = 0.45$ V when C = 1 |
| | 1 | Low at $V = 0.45$ V when C = 0 <br><br> High at $V = 0.45$ V when C = 1 |

Table 7: Truth table for Toffoli gate

| B | $C_1$ | $C_2$ | Toffoli |
|---|---|---|---|
| 0 | 0 | 0 | 0 |
| 0 | 0 | 1 | 0 |
| 0 | 1 | 0 | 0 |
| 0 | 1 | 1 | 1 |
| 1 | 0 | 0 | 1 |
| 1 | 0 | 1 | 1 |
| 1 | 1 | 0 | 1 |
| 1 | 1 | 1 | 0 |

Table 8: Assignments of logic values for the implementation of Toffoli gate

| | Logic value | Current |
|---|---|---|
| B | 0 | High at $V = 0.45$ V |
| | 1 | Low at $V = 0.45$ V |
| $C_1$, $C_2$ | 0 | Low at $V = 0.9$ V |
| | 1 | High at $V = 0.9$ V |
| Toffoli | 0 | High at $V = 0.45$ V when $I(0.9$ V) is low |
| | | Low at $V = 0.45$ V when $I(0.9$ V) is high |
| | 1 | Low at $V = 0.45$ V when $I(0.9$ V) is low |
| | | High at $V = 0.45$ V when $I(0.9$ V) is high |

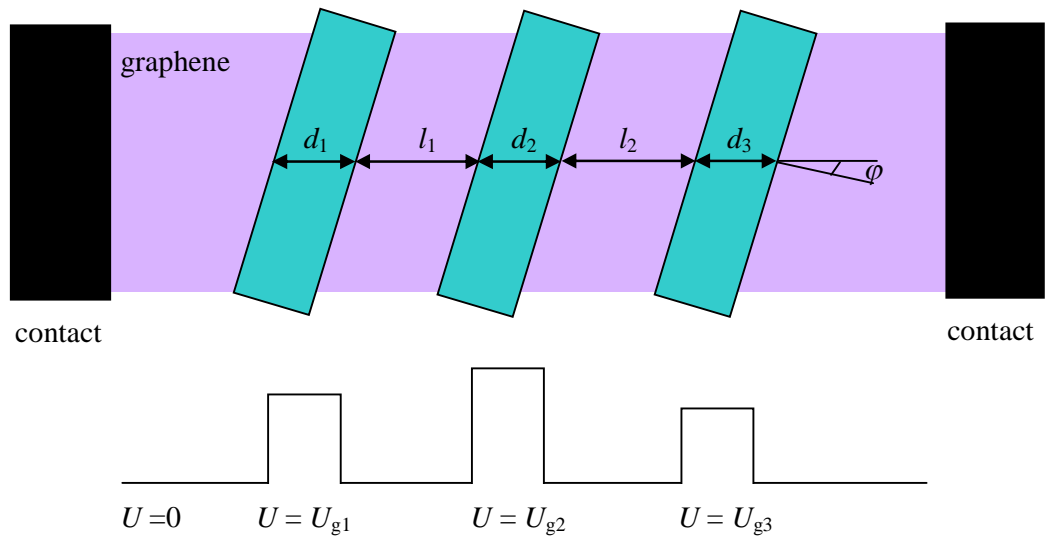

Fig.1  Schematic representation of a device that can implement three-input reversible logic

gates in graphene (top) and the associated potential energy (bottom)

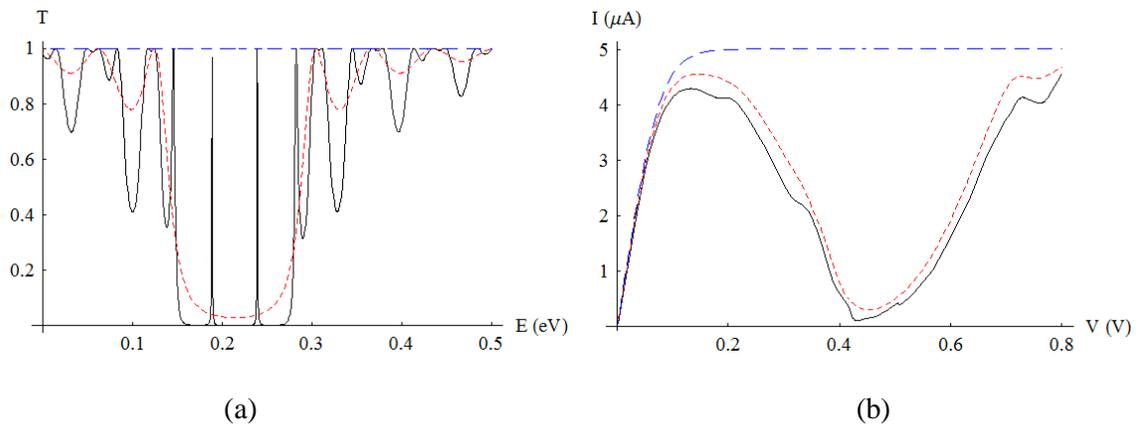

(a)　　　　　　　　　　　(b)

Fig. 2 (a) Transmission and (b) current through the device in Fig. 1 when $U_{g1} = U_{g2} = U_{g3} = 0$ (blue dashed line), when $U_{g1} = 0.2$ eV, $U_{g2} = U_{g3} = 0$ or $U_{g2} = 0.2$ eV, $U_{g1} = U_{g3} = 0$ (red dotted line), and when $U_{g1} = U_{g2} = 0.2$ eV, $U_{g3} = 0$ (black solid line).

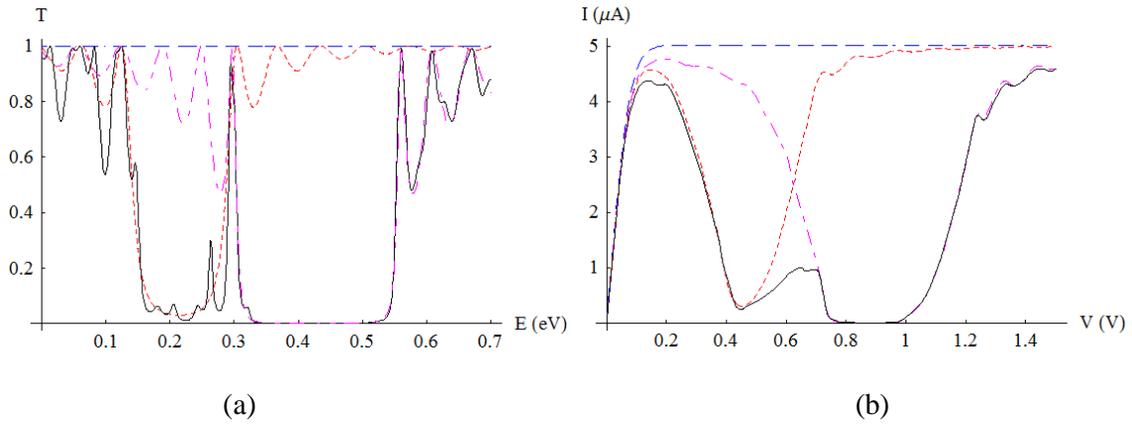

(a)                                          (b)

Fig. 3  (a) Transmission and (b) current through the device in Fig. 1 when $U_{g1} = U_{g2} = U_{g3} = 0$ (blue dashed line), when $U_{g1} = 0.2$ eV, $U_{g2} = U_{g3} = 0$ (red dotted line), when $U_{g2} = 0.3$ eV, $U_{g1} = U_{g3} = 0$ (magenta dashed-dotted line) and when $U_{g1} = 0.2$ eV, $U_{g2} = 0.3$ eV, $U_{g3} = 0$ (black solid line).

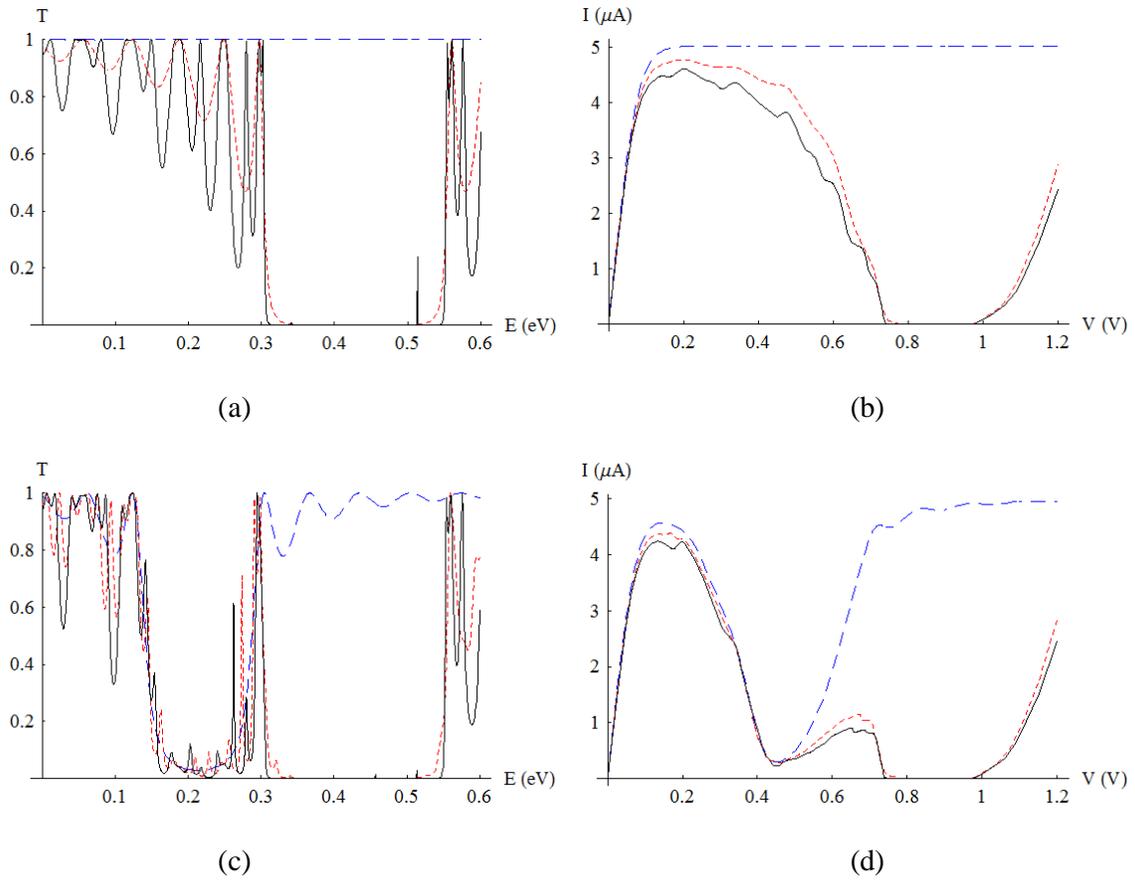

(a)  (b)

(c)  (d)

Fig. 4 (a) Transmission and (b) current through the device in Fig. 1 when $U_{g1} = 0$ and $U_{g2} = U_{g3}$ = 0 (blue dashed line), $U_{g2} = 0.4$ eV, $U_{g3} = 0$ or $U_{g2} = 0$, $U_{g3} = 0.4$ eV (red dotted line), and when $U_{g2} = U_{g3} = 0.4$ eV (black solid line). (c) and (d): similar to (a) and (b) but for $U_{g1} = 0.2$ eV